\documentclass[conference]{IEEEtran}

\usepackage{color, colortbl}
\usepackage{multirow}
\usepackage{subfigure}
\usepackage{transparent}
\usepackage{balance}
\usepackage{url}
\usepackage{graphicx}
\usepackage{paralist}
\usepackage{amsmath, amssymb, amsthm, framed, graphics}
\usepackage{bytefield}
\usepackage{todonotes}
\usepackage{algorithm}
\usepackage{algpseudocode}
\usepackage[group-separator={,}]{siunitx}
\usepackage[underline=true]{pgf-umlsd}
\usetikzlibrary{calc}
\let\labelindent\relax
\usepackage{enumitem}
\usepackage{pifont}

\setenumerate{noitemsep,topsep=0pt,parsep=0pt,partopsep=0pt}
\setitemize{noitemsep,topsep=0pt,parsep=0pt,partopsep=0pt}

\usepackage{framed}

\newtheorem{theorem}{Theorem}

\definecolor{LightCyan}{rgb}{0.88, 1, 1}
\definecolor{LightRed}{rgb}{1, 0.88, 0.88}
\definecolor{LightGreen}{rgb}{0.88, 1, 1}
\definecolor{LightYellow}{rgb}{1, 1, 0.88}
\definecolor{LightOrange}{rgb}{1, 0.92, 0.77}

\newcommand{\ccnname}[1]{{\footnotesize\texttt{#1}}}
\newcommand{\Adv}{Adv}
\newcommand{\Eve}{$\mathcal{E}$}
\newcommand{\Evem}{\mathcal{E}}
\newcommand{\content}{{\em content}}
\newcommand{\interest}{{\em interest}}

\newcommand{\pint}{{\em pInt}}
\newcommand{\pints}{{\em pInt} messages}
\newcommand{\cdata}{$\mathsf{CrSD}$}
\newcommand{\cdatas}{$\mathsf{CrSD}$-s}
\newcommand{\cdatam}{\mathsf{CrSD}}
\mathchardef\mhyphen="2D
\newcommand{\secdata}{$\mathsf{Sec}\mhyphen\mathsf{CrSD}$}
\newcommand{\secdatam}{\mathsf{Sec}\mhyphen\mathsf{CrSD}}
\newcommand{\adata}{$\mathsf{A}\mhyphen\mathsf{CrSD}$}
\newcommand{\adatas}{$\mathsf{A}\mhyphen\mathsf{CrSD}$-s}
\newcommand{\adatam}{\mathsf{A}\mhyphen\mathsf{CrSD}}
\newcommand{\ignore}[1]{}

\newtheorem{definition}{Definition}

\begin{document}
\title{Practical Accounting in Content-Centric Networking (extended version)}
\vspace{-1em}
\author{\IEEEauthorblockN{Cesar Ghali \quad\quad Gene Tsudik \quad\quad Christopher A. Wood}
\IEEEauthorblockA{University of California, Irvine\\
\{cghali, gene.tsudik, woodc1\}@uci.edu}
\and
\IEEEauthorblockN{Edmund Yeh}
\IEEEauthorblockA{Northeastern University\\
eyeh@ece.neu.edu}}

\maketitle

\begin{abstract}
Content-Centric Networking (CCN) is a new class of network architectures designed to 
address some key limitations of the current IP-based Internet. One of its main features 
is in-network content caching, which allows requests for content to be served by 
routers. Despite improved bandwidth utilization and lower latency for 
popular content retrieval, in-network content caching offers producers
no means of collecting information about content that is requested and later 
served from network caches. Such information is often needed for accounting purposes. 
In this paper, we design some secure accounting schemes that vary in the 
degree of consumer, router, and producer involvement. Next, we identify and analyze 
performance and security tradeoffs, and show that specific per-consumer accounting
is impossible in the presence of router caches and without application-specific support. We 
then recommend accounting strategies that entail a few simple requirements for CCN architectures.
Finally, our experimental results show that forms of \emph{native and secure} 
CCN accounting are both more viable and practical than application-specific 
approaches with little modification to the existing architecture and protocol. 
\end{abstract}

\section{Introduction}
The original Internet was designed in the late 1970s with the main purpose of providing end-to-end 
communication. It allowed thousands of users to remotely access scarce computing resources from 
terminals.  Since then, the number of Internet users has grown exponentially.  
They use a wide variety of applications, many of which involve some form of content distribution.  
This shift in usage exposed some design limitations of the current IP-based Internet and motivated
the exploration of new networking architectures.

Content-Centric Networking (CCN) is an approach to inter-networking exemplified by two well-known 
research efforts: the CCNx project at the Palo Alto Research Center~\cite{content-centric} and 
Named-Data Networking (NDN)~\cite{jacobson2009networking}. In IP-based networking, a user
requests \content\ by addressing the host at which it is stored. Conversely, in CCN, \content\ is 
assigned a unique name and is addressed directly. Any entity can become a content {\em producer} as long
as it can show that is authorized for a certain part of the global content namespace.
A user, called a consumer, requests \content\ by issuing an \interest\ carrying the former's 
name. Such interests can be satisfied by any entity (host or router) that either creates or 
caches the requested content. The \content\ follows, in reverse, the exact path of the preceding 
interest towards the consumer. Any intervening routers on this path may cache the content to 
satisfy future interests for the same content.

These in-network caches facilitate efficient content distribution.
This important feature helps reduce end-to-end latency and lower bandwidth consumption when requesting 
popular content. Since an interest can
be satisfied by a copy of the requested content found in a router's cache, it might not reach the 
producer. Consequently, a producer might only receive a small 
fraction of all interests for a given piece of content. At the same time, the number, sources, and timing 
of interests represent important information that could be used for accounting by the producer. 
Even if the timing and the number of interests were somehow communicated to the producer, interest sources
would remain unknown since CCN lacks consumer information, e.g., source addresses, in interests. 

Furthermore, router cache space will likely be treated as a valuable (and even 
premium) resource as CCN is deployed in the real world. Thus, a mechanism is 
needed for reporting cache hits 
to content producers or router owners, thereby informing them about content usage. 
To be viable, such a mechanism must only incur minimal bandwidth, 
computation, and storage overheads. Finally, to prevent attacks such as false 
cache usage reporting, it also must be be secure. 
In this paper, we design a lightweight secure accounting mechanism, applicable, 
to both CCNx and NDN\footnote{Support for NDN requires minor packet format and 
protocol changes.}. Our intended contributions are three-fold:
\begin{itemize}
\item Identification and motivation for features needed for CCN accounting and for security thereof.
\item The first comprehensive technique for content and cache usage accounting,
with varying levels of consumer, router, and producer involvement.
\item Analysis of performance and security tradeoffs.
\end{itemize}
In the rest of this paper, we use the term CCN to refer to both CCNx and NDN.

\noindent
{\bf Organization.} Section \ref{sec:overview} overviews CCN. Next, Section \ref{sec:open-env} 
discusses desired features for content accounting. Security requirements are addressed 
in Section \ref{sec:security}. Performance of the proposed approach is assessed in Section \ref{sec:experiments}. 
The paper concludes with a summary of related work in Section \ref{sec:related}.

\section{CCN Overview}
\label{sec:overview}
This section gives an overview of CCN. Given familiarity with CCN, it can be skipped without loss of continuity. 
Note that all details are presented in the context of the CCNx architecture and protocol. Specifics such as packet 
formats, message fields, and routing decisions have subtle differences in NDN. However, with minor
changes to the protocol and packet formats, our description also applies to NDN.

Unlike IP which focuses on end-points of communication and their 
names/addresses, CCN stresses content by making it named, addressable, and routable~\cite{content-centric}. 
A content name is represented with a labeled content identifier (LCI) string, which
is similar to a standard URI. For example, a BBC news home-page for April 15, 2015 might 
be named \ccnname{lci:/uk/bbc/news/2015APR15/index.htm}.

CCN communication follows the {\em pull} model whereby content is delivered to consumers only upon 
explicit request. There are two basic types of packets (messages) in CCN: interest and content object 
messages\footnote{We use the terms \content\ and content object interchangeably.}. A consumer 
requests content by issuing an \interest\ message. If an entity can ``satisfy'' a given interest, it returns the 
corresponding \emph{content object}. Name matching in CCN is exact, i.e., 
an interest for \ccnname{lci:/snapchat/Bob/video-749.avi} can only be satisfied 
by a content object named \ccnname{lci:/snapchat/Bob/video-749.avi}.

CCN interest messages include, at a minimum, the name of the requested content. They may also
carry a payload that enables consumers to push data to producers along with the content 
request~\cite{mosko2013ccnx}.\footnote{Currently, NDN interest messages do not provide an arbitrary-length 
payload and therefore cannot support some of the proposed accounting types, as will become clear 
later in the paper. However, if in the future the NDN interest format includes a field similar to the CCNx 
payload, our accounting approach will become applicable.} CCN content objects 
include a number of fields. In the context of this paper, we are only interested in the following four:
\begin{itemize}
\item \texttt{Name} -- A URI-compliant sequence of name components.
\item \texttt{Payload} -- The actual data of the content object.
\item \texttt{Validation} -- A composite of validation algorithm information 
(e.g., the signature algorithm used, its parameters, and a link to the public 
verification key), and validation payload (e.g., the signature). For simplicity, 
we use the term ``signature'' to refer to this field.
\item \texttt{ExpiryTime} -- Producer-recommended time for the content objects to be cached.
\end{itemize}
There are three types of CCN entities or roles:\footnote{A physical entity (a host, in today's parlance) can be 
both consumer and producer of content.}
\begin{itemize}
\item {\em Consumer} -- an entity that issues an interest for content.
\item {\em Producer} -- an entity that produces and publishes content. 
\item {\em Router (Forwarder)} -- an entity that routes interest packets and forwards corresponding content packets. 
\end{itemize}
Each CCN entity maintains three structures~\cite{CCNxNodeImplementation}:
\begin{itemize}
\item {\em Content Store} (CS) -- a cache used for content caching and retrieval. Cache size is determined by 
local resource availability. Each router may unilaterally determine whether to cache content and for how long. 
From here on, we use the terms {\em CS} and {\em cache} interchangeably. 
\item {\em Forwarding Interest Base} (FIB) -- a table of name prefixes and corresponding outgoing interfaces. 
The FIB is used to route interests based on longest-prefix-match of the name. 
\item {\em Pending Interest Table} (PIT) -- a table of outstanding (pending) interests and a set of 
corresponding incoming interfaces.\footnote{A given PIT entry might indicate multiple incoming interfaces
reflecting the possibility of multiple interests arriving at, or near, the same time.}
\end{itemize}
Upon receiving an interest with the name $N$ (i.e., for the content with name $N$), 
a router first checks its cache for existence of a local copy of  
content with the same name.\footnote{This is why CCN lacks any notion of a {\em destination address}; 
since content can be served by any CCN entity.} 
If a local copy is not found in the cache, and there are no pending interests for $N$, the router forwards 
the interest to the next hop(s) according to its FIB and forwarding strategy. 
For each forwarded interest, a router creates a new PIT entry
with state information, including the name and the interface on which it arrived. Moreover, if an 
interest with the name $N$ arrives while there is already an entry for the same name in the 
PIT, the router collapses the present interest and only stores the interface on which it was received. 
When content is returned, the router forwards it out on all recorded incoming interfaces and flushes the 
corresponding PIT entry. Since no additional information is needed to deliver content, an interest does not 
carry a source address.
\section{Accounting in CCN} \label{sec:open-env}
As mentioned earlier, router caches present a major challenge for accounting in CCN.
In particular, if interests are satisfied by caches, {\bf how can a content producer collect information 
about the popularity of (or demand for) its content}? 
In this section, we discuss design elements for accounting in CCN.  
For the time being, we do not emphasize security and privacy considerations. 
Specifically, assuming benign (well-behaved) consumers, routers, and producers, our initial 
goal is to determine the minimal functionality needed by all CCN entities to facilitate {\em correct} accounting. 
In doing so, we consider three types of accounting information:
\begin{compactitem}
\item {\em Individual:} This type of information is tied directly to a specific consumer. An example 
might be the number of times a particular consumer requested a particular content. It 
provides linkability between consumers and content they obtain. Moreover, it requires 
revealing consumer identities, at least to the producer.
\item {\em Distinct:} This type of information is functionally equivalent to the individual
accounting information with the exception that consumer identities are not revealed.
Instead, a randomly generated nonce is added to the each interest to enable
a producer to distinguish between separate interests. More details are explained in 
the following sections.
\item {\em Aggregate:} This type of information represents an aggregate over a set of consumers. 
For example, it might include the number of times a particular piece of content was requested 
from a specific geographic location or an ISP. Aggregate information enables some degree of consumer privacy. 
\end{compactitem}
We believe that these three types are sufficiently representative of any accounting information 
needed in any real-world CCN application and focus on them in the remainder of this work.
Also, we claim that accounting should not be mandatory for all content. Some producers might
not care about the popularity of {\em any} of their content, while others might need accounting information
only for {\em some} of their content. We refer to content for which producers desire such 
information as {\em accountable} content.

Lastly, one important design dimension is whether accounting information is reported in real time (online)
or not (offline). In the latter case, a network management protocol can be envisaged 
whereby an AS-level accounting server periodically collects cache hit logs from its routers
and reports the results to individual producers or AS-housing producers.
This kind of accounting seems viable. However, it involves a potentially significant delay
in notifying producers about the demand, or lack thereof, for their content. This
might be unacceptable for content for which real time demand information is needed.
Intuitively, real-time accounting is a more difficult problem to solve. 
We therefore limit the rest of this paper to real-time accounting.

\subsection{Counting Cache Hits vs. Content Requests}
\label{sec:accounting-types}
Another variable in supporting CCN accounting is exactly what is being counted: instances
of cache hits, or instances of requested content being served to the consumer? 
A cache hit occurs when a router or another entity finds the requested content in its cache.
By {\em another entity} we mean either a producer or a repository that 
keeps a copy of the content. We assume that accounting for cache hits is only relevant for routers,
i.e., network elements.\footnote{Accounting for cache hits in repositories or at producers themselves
is out of scope.} An instance of content being served occurs when a cache hit takes 
place {\bf and} the content is actually delivered to a \emph{single} consumer.

It might seem that these two types of events are the same, i.e., a content is served once
for every cache hit, and vice-versa. However, this is not the case in CCN.
Whenever a router receives an interest, it may choose 
to multicast (forward) it out on multiple interfaces. This behavior is officially 
allowed since a router's FIB can express multiple next hops for a given name prefix.
One practical reason for allowing it that it facilitates quick(er) fetching of content.
However, it also complicates accounting. Consider the following scenario:  
\begin{quote}
Suppose that a consumer requests content $CO$ and the issued interest is received by router $R_1$.
The latter then forwards the interest for $CO$ to two upstream routers $R_2$ and $R_3$
based on its FIB.  Both $R_2$ and $R_3$ have $CO$ in their respective caches and each replies to 
$R_1$ with its cached version. Assuming that $R_2$'s copy of $CO$ is the first to reach $R_1$, the latter 
forwards $CO$ downstream and flushes the appropriate PIT entry. When $R_3$'s copy of $CO$ arrives,
$R_1$ discards it since it does not refer to a current existing PIT entry. If both $R_2$ and $R_3$ 
inform $CO$'s producer $P$ about a cache hit on $CO$, $P$ would incorrectly assume that $CO$
was requested twice. Even though, technically, $CO$ is served twice by two distinct
routers, there was only one requesting consumer which received only one copy of $CO$.
\end{quote}
According to this scenario, the count of cache hits can exceed the content request count. 
This problem occurs because there is no way to distinguish among different interests issued for the 
same content.\footnote{NDN interests carry a random nonce used for interest loop detection, 
which can be helpful in this distinction. However, CCNx interests do not carry nonces.} 
In other words, if consumers $Cr_1$ and $Cr_2$ issue interests 
for $CO$vat different times, their interests would be identical. Moreover, even if $CO$ is not cached, i.e.,
interests for it reach $P$, and if $Cr_1$ and $Cr_2$ issue interests for $CO$ at roughly the same 
time, $P$ would be unable to distinguish between this case -- when two consumers ask for $CO$ -- and 
the case in the scenario above -- when one consumer asking for $CO$ and $R_1$ decides to multicast the 
interest upstream. Note that the number of cache hits is equal to two in both cases, but the number of 
content objects served is two and one, respectively.

The reason for supporting both types of accounting is quite intuitive: a producer might need to 
know the exact demand for its content, whether on an aggregate or individual basis.
Separately, a producer might need to know which routers experience cache 
hits for its content. The latter could be used to reconcile billing the producer for cache usage. 

Finally, even though accounting for cache hits and content requests is not the same thing,
we naturally would like to use the same mechanism as much as possible to provide both. 
Therefore, for the rest of the paper and unless otherwise mentioned, we use the term {\em accounting} 
to refer to both accounting for cache hits and content requests.

\subsection{Accounting via Content Encryption}
One intuitive accounting approach is to use encrypted content.\footnote{This is also 
a form of access control.} Suppose that producers encrypt all accountable content, 
and the decryption keys -- which, in CCN, are represented as content objects with well-defined names -- 
are configured not to be cached, i.e., by setting their {\tt ExpiryTime} to $0$. Even if consumer interests requesting such 
content are satisfied from in-network caches, the former must separately issue interests requesting the 
decryption key(s). Such interests bypass in-network caches and reach the producers, thereby 
enabling per-request accounting.

With content encryption, the desired type of accounting dictates 
how interests requesting keys should be generated. 
In the case of individual accounting, consumers must include some kind of consumer-specific data 
in the interests when keys are requested. Such data allows producers to link these interests to 
specific consumers. However, if only aggregate accounting is required, interests requesting 
keys do not have to carry any consumer-specific data. As mentioned above, such interests need
to have some kind of a nonce to enable the producer to distinguish between the case of receiving two 
interests from two different consumers, or receiving two interests sent from a single consumer 
and were multicasted by a router in the network.


Accounting via content encryption has two primary advantages: (1) it is transparent to the network layer, 
and (2) it does not require any new features and message types. However, despite
its apparent simplicity, it is not efficient. All accountable content objects need 
to be encrypted and keys need to be requested and distributed
separately. Thus, content is obtained by issuing two interests  -- one for the content and 
one for the key(s) -- thus incurring at least two round-trips to the producer.

We believe that an ideal accounting mechanism should efficiently work for all accountable 
content. That is, it should not require a consumer to issue more than a single interest for
accountable content. Also, general content accounting should be distinct from content access control.

\subsection{Accounting via Push Interests} \label{sec:pints}
The accounting approach proposed in this paper is based on real-time reporting. 
The key element is a new message type that we call a \emph{push} interest, denoted 
as \pint. Its main purpose is to inform the producer that its content has been requested and a cache hit occurred. 
Structurally, a \pint\ carries a name similar to a regular interest. However, the most important
distinguishing feature of a \pint, is that, unlike a regular interest, it 
does not leave behind any state in routers. Specifically, a \pint\ referencing content 
$CO$ is forwarded by each router until it reaches its corresponding producer $P$, and no information
about that \pint\ is retained by any intervening router. A router forwards a \pint\ just like it forwards
a regular interest with the exception that \pints\ are not multicasted. 
This restriction is necessary to prevent producers from receiving duplicate copies of the same \pint. 

Besides this forwarding change, the behavior of CCN routers is slightly modified to support \pint\ generation. 
There are two cases when a router generates a \pint: 
\begin{compactenum}
\item Whenever a regular interest with name $N$ is satisfied from its cache, a router generates a 
\pint\ with the same name $N$ and forwards it upstream towards the producer. 
\item When a router receives a content object corresponding to a PIT entry, 
it forwards that message on all downstream interfaces listed in said PIT entry.
However, before flushing that entry, a router generates a \pint\ 
that aggregates all collapsed interests. (These aggregation 
details are discussed in Section~\ref{sec:pints-format}.) Note that {\em collapsed}
refers only to those interests that were not originally forwarded upstream. This is because 
the one forwarded upstream presumably already (1) reached the producer, or (2) triggered its own \pint\ via 
case 1 above.\footnote{For example, suppose that a router receives an interest for content $CO$ on interfaces: 
2, 3, 5, and 6. Regularly, only the first one (arriving on interface 2) is forwarded, say, on interface 9. 
Interests on 3, 5 and 6 arrive later and are collapsed in the same PIT entry. Now, when $CO$ arrives
back on interface 9, it is forwarded out on all 4 incoming interfaces. However, the router generates 
a \pint\ that reflects only the last 3 interfaces: 3, 5 and 6.}
\end{compactenum}
%
%

In order for a producer to inform routers about what content requires accounting, we also introduce a new 
{\tt ACCT} flag in the content header, which reflects one the following three values:
\begin{compactenum}
\item \texttt{NONE}: the producer requests no accounting information for this content.
\item \texttt{AGGREGATE}: the producer requests aggregate accounting information for this content.
\item \texttt{DISTINCT}: the producer requests distinct accounting information for this content. 
\item \texttt{INDIVIDUAL}: the producer requires individual interest-level accounting for this content. 
\end{compactenum}
Whenever a cache hit occurs, routers behave the same for cases 2, 3, and 4. 
The only difference is when a content arrives and a router has a number of 
previously collapsed interests for that content. In case 2, a router generates a \pint\ with 
the count of collapsed interests for a given content. In cases 3 and 4, a router reports 
the {\em actual} interests, which can optionally be bundled into a single \pint.

As hinted in the above discussion, we aim to support aggregate, distinct, and individual 
accounting types, and to report instances of cache hits as well as instances of content requests.
Distinct and individual accounting for cache hits and content requests (as well as aggregate 
for cache hits) is possible and indeed attained by the proposed technique. 
Unfortunately, as will become clear 
below, supporting accurate distinct and aggregate accounting for {\em content requests} is quite challenging. 
However, if some kind of consumer-specific data is provided in interest {\tt Payload} fields, 
probabilistically accurate distinct and aggregate accounting for content requests can be achieved.

\subsection{\pint\ Format and Features} \label{sec:pints-format}
We now describe the \pint\ message format and discuss the purpose of its fields.
Structurally, this message is nearly isomorphic to CCNx 1.0 interests~\cite{CCNxMessagesTLVFormat},
and includes the following necessary fields:
\begin{compactitem}
\item {\tt Name}: copied entirely from the {\tt Name} field in the interest (or PIT entry) 
that triggers a \pint.
\item {\tt Type}: flag indicating whether this \pint\ is for aggregate, distinct, or 
individual accounting.
\item {\tt Origin}: identifies the router that generates the \pint, e.g., the router's prefix 
(if available) or public key digest.
\item {\tt Count}: set to 1 in the case of a cache hit, or the number of interfaces minus one 
on which the content object was forwarded downstream if interest collapsing occurred.
\item {\tt Cdata}: a random nonce or consumer-specific data used by producer for different 
purposes based on the accounting type required (i.e., distinct or individual). If {\tt Count}$>1$,
this is a sequence of {\tt Count} of consumer-specific data culled from corresponding interests. 
Such data can be carried in the interest {\tt Payload} field.
\end{compactitem}
The semantics of the {\tt Cdata} field depend on the type of required accounting 
information. As stated above, aggregate accounting for cache hits does not require 
{\tt Cdata} to be present. In the following we discuss consumer-specific data 
requirements for other accounting types.

\noindent
{\bf Aggregate accounting for content requests:} The problem in this type of accounting 
is that producers do not have the means to distinguish between the cases where received
interests (or \pints) with the same name are multicasted by routers or generated by 
several distinct consumers. However, if consumers include random nonces and timestamps as 
consumer-specific data, this distinction can be achieved.

\noindent
{\bf Individual Accounting for cache hits and content requests:} In this case, {\tt Cdata} 
needs to reflect the identify of consumers issuing interests. This value can take a variety of forms:
\begin{enumerate}
\item Consumer public keys or their digests. Note that this form reveals consumer identities 
to all network entities -- not only producers.
\item Group public keys or their digests. A group can be an organization, 
autonomous system (AS), or a geographical region. In this case, the group identify is revealed 
rather than that of the individual consumer.
\item Unique consumer identifiers (i.e., pseudonyms). Although this does not violate 
consumer anonymity, such identifiers need to be assigned to consumers by producers or a trusted 
third party before any interests are issued. Note that this form of {\tt Cdata} allows 
interest linkability.\footnote{Interest linkability is defined as the ability of an 
eavesdropper (observer or adversary) to reveal that fact whether two captured interests are 
issued by the same consumer.}
\item Consumer identity (using any of the three previous forms) with nonces and timestamps. 
This form of consumer-specific data allows producers to know which consumers request what 
content, as well as how many times such requests are made.
\end{enumerate}

\noindent
{\bf Distinct Accounting for cache hits and content requests:} For this accounting type, 
{\tt Cdata} needs to reflect the uniqueness of interests. This can be achieved if consumers 
include a nonce and timestamp in the {\tt Cdata} field. The format of the nonce is application-specific 
and can range from a random number to the hash of the content name and the timestamp.\\
Note that the same knowledge provided to the producer in the distinct accounting 
case can also be attained using aggregate accounting type if {\tt Cdata} reflects the uniqueness 
of interests. However, we keep the distinction between these two types for ease of 
classification.

Each of the above forms impose different overhead on consumers and producers. 
However, router overhead is only very slightly affected if accounting is done with \pints.\footnote{
Content encryption-based accounting is completely transparent to routers.} This
is because routers simply populate the {\tt Cdata} field of generated \pints\ using 
information contained in the {\tt Payload} field of the corresponding interests, regardless 
of how consumer-specific data is generated. In other words, \emph{routers are oblivious to 
the accounting type used}.

Also, note that the choice of which form to use is an application-specific issue. 
We do not mandate a specific technique. 

\subsection{Correctness}
We define correctness of an accounting technique as follows.

\begin{definition}
\label{def:correctness}
An accounting technique is {\em correct} if it accurately
reports cache hit and content request information to the producer, assuming that 
all participants faithfully follow the rules (i.e., exhibit no malicious behavior) and
there are no transmission errors, packet loss, or node failures that affect  
accounting-relevant traffic. 
\end{definition}

We also define probabilistically correct accounting technique as follows.
\begin{definition}
An accounting technique is {\em probabilistically correct} if it is correct 
with a negligible probability of error, i.e., reporting inaccurate or false information.
\end{definition}

We now informally prove correctness of each of the proposed accounting techniques: 
individual, distinct, and aggregate in both cache hits and content requests cases.

\noindent
{\bf Cache Hits:} A router $R$ generates 
\pints\ for producer $P$ for every cache hit on accountable content objects.
Since all routers (including $R$) 
on the path to $P$ forward \pints\ according to their FIB entries, such messages are 
guaranteed to be received by $P$.\footnote{Recall that we assume there are no transmission
errors or packet loss events.} This provides accurate cache hit counts for content 
objects in question. This argument holds for all three types of accounting. The only 
difference is that {\tt Cdata} fields of \pint\ must contain appropriate consumer-specific 
data in some accounting types.

\noindent
{\bf Content Requests:} 
Recall that \pints\ are also generated to report collapsed 
interests. Such interests reflect content requests in the network. 
Although \pints\ 
provide {\em correct} individual, distinct, and aggregate accounting for cache hits, it can 
only provide {\em probabilistically correct} accounting for content requests. 
As stated above, consumers can include nonces and timestamps in the {\tt Payload} 
fields of their interests. If producers post-process received \pints, this 
{\tt Payload} information allows them to distinguish between the cases where received interests 
(or \pints) with the same name are multicast by routers or generated 
by several distinct consumers. However, since nonces are random strings of bits, collisions 
might occur with a probability negligible in their bit length.

\subsection{Performance} \label{sec:performance}
As previously mentioned, our goal is to design a lightweight mechanism for secure accounting 
in CCN. In this section we discuss involvement of consumers, routers and producers.

\noindent
{\bf Consumers:} The degree of consumer overhead varies depending on the 
accounting type required. In the case of aggregate accounting for cache hits, \pint-based accounting 
is transparent to the consumers. However, when aggregate accounting for content requests
is required, distinct or individual accounting techniques must be used. In this case, consumers
must include a random nonce or consumer-specific data in all issued interests as described 
in Section~\ref{sec:pints}.\footnote{Human consumers are 
not involved, only their applications.} The 
overhead for consumers depends on how this data is generated. 
For instance, consumer-specific data resilient to some types of attacks can increase 
this overhead -- see Section~\ref{sec:mitigating-replay} for details.

\noindent
{\bf Routers:} Regardless of the accounting type used, router involvement is always 
required when \pints\ are used.\footnote{Content encryption-based accounting is 
transparent to routers.} For all accountable content, routers must generate \pints\ whenever 
cache hits occur or collapsed interests are satisfied. However, we argue that routers overhead 
is minimal. 

\begin{algorithm}[t]
\caption{{\sf \pint-Generation}} \label{alg:pint-generation}
\small
\begin{algorithmic}[1]
\State {\bf Input:} $CO[N,~${\tt ACCT}$]$, $Int[N,PL]$, $R_{id}$
\State \pint.{\tt Name} := $CO.N$
\State \pint.{\tt Type} := $CO.${\tt ACCT}
\State \pint.{\tt Origin} := $R_{id}$
\If {$CO$ from local cache}
	\State \pint.{\tt Cdata} := $Int.PL$
	\State \pint.{\tt Count} := $1$
\Else
	\State $e$ = {\sf FindPITEntry}($CO.N$)
	\For {each $i$ in $e/\{Int$\}}
		\State \pint.{\tt Cdata} := \pint.{\tt Cdata} $||$ $i.PL$
		\State \pint.{\tt Count} := \pint.{\tt Count} + $1$
	\EndFor
\EndIf
\State Forward \pint\ according to the FIB
\end{algorithmic}
\end{algorithm}

\begin{table*}[t!]
\begin{center}
\caption{CCN entities overhead in accounting. N/A also indicates that no involvement is required.}
\label{tbl:performance}
\small
\begin{tabular}{|c|c|c|c|c|c|c|} \hline
& \multicolumn{2}{c|}{\bf Consumers} & \multicolumn{2}{c|}{\bf Routers} & \multicolumn{2}{c|}{\bf Producers} \\ \hline
{\bf Accounting Type} & {\bf Aggregate} & {\bf Individual} & {\bf Aggregate} & {\bf Individual} & {\bf Aggregate} & {\bf Individual} \\ \hline
{\bf Cache hits} & N/A & consumer-specific & \multicolumn{2}{c|}{\multirow{2}{*}{\pints generation}} & $\mathcal{O}(|\mathbb{C}_P|)$ & $\mathcal{O}(|\mathbb{U}_P| \cdot |\mathbb{C}_P|)$ counters \& \\ \cline{1-2}
\bf Content requests & nonces/timestamps & data generation & \multicolumn{2}{c|}{} & counters & consumer-specific data sanitation \\ \hline
\end{tabular}
\end{center}
\end{table*}

To enable efficient throughput, a router must be able to quickly construct \pints\ 
in the fast path. All \pint\ field values can be copied 
from their corresponding counterparts in interest and content headers. For instance, the
{\tt Name} and {\tt Type} values can be copied from the corresponding content header fields {\tt Name} 
and {\tt ACCT}. The {\tt Cdata} field can be populated by copying {\tt Payload} field values from 
corresponding interests. Moreover, the router identifier in the {\tt Origin} field can be pre-computed 
(if needed) and stored in memory. 

Algorithm~\ref{alg:pint-generation} illustrates a procedure to generate \pints\ in
routers. It is triggered whenever a router receives a solicited content object or locates a copy 
in its local cache. The algorithm takes as input the received content object $CO$ with name $N$ 
and accounting flag {\tt ACCT}, its corresponding interest $Int$ with name $N$ and payload $PL$, and 
the router $R_{id}$. If $CO$ is not in local cache, the router (1) copies consumer-specific data 
from all interests in the corresponding PIT entry into {\tt Cdata}, and (2) sets {\tt Count} value 
accordingly.

To obtain better better bandwidth utilization, routers can generate a single \pint\ 
message for a specific content $CO$ in a pre-defined time window. In this case, routers 
report \emph{batched} cache hits by (1) including the actual cache hit counters in the \pint's 
{\tt Count} field, and (2) listing all consumer-specific data from the corresponding interests 
(if any) in the \pint's {\tt Cdata} field. 

\noindent
{\bf Producer:} Similar to routers, producers requesting accounting information are always 
involved and the overhead varies depending on the type of accounting required. We describe 
these differences below.
\begin{itemize}
\item Aggregate: the overhead is minimal in this case. Producers only need to maintain a 
cache hit counter for every accountable content they publish. 
The counters maintained  
at producer $P$ are in order $\mathcal{O}(|\mathbb{C}_P|)$, where $\mathbb{C}_P$ is the set 
of accountable content published by $P$.
\item Individual: producers have to maintain a counter for every \{consumer, 
accountable content\} pair. 
The size of the counters table is in order 
$\mathcal{O}(|\mathbb{U}_P| \cdot |\mathbb{C}_P|)$, where $\mathbb{U}_P$ is the set of 
consumers requesting accountable content from $P$. The producer might also store 
the timestamp at which requests for content arrived. This, obviously, storage requirements. 
Moreover, to provide {\em correct} individual accounting for content requests, 
producers must post-process all received \pints\ 
in order to detect duplicates caused by interest multicasting by routers.
\item Distinct: in this case, producers have to maintain a counter for every accountable 
content object as well as information about all requests for these objects. Such 
information could be the timestamp at which requests are received, as discussed
in Section \ref{sec:pints}. The size of the 
stored information is in order $\mathcal{O}(|\mathbb{R}_P| + |\mathbb{C}_P|)$, 
where $\mathbb{R}_P$ is the set of all requests received by $P$ for its published 
accountable content.
\end{itemize}
Table~\ref{tbl:performance} summarizes each CCN entity overhead in the proposed accounting 
mechanisms.
\section{Security Considerations \label{sec:security}}
Thus far, we assumed that all entities involved in 
accounting are benign, i.e., act honestly and correctly. However, this assumption is clearly unrealistic in 
practice. In this section, we propose security requirements to achieve secure 
accounting in CCN. We also show that some attacks cannot be prevented or 
even detected without additional cryptographic overhead. Moreover, we stress that 
implementing secure accounting incurs a trade-off between security and overhead 
on consumers and producers. Fortunately, routers are unaffected by such amendments. 

\subsection{Adversary Model}
The anticipated adversary \Adv\ is a malicious router generating \pints\ 
for non-existing interests, when individual accounting is required. 
In other words, \Adv\ tries to inflate 
individual accounting information for both cache hits and content requests. 
We also assume that consumers behave honestly: if a consumer 
needs to provide consumer-specific data or a random nonce, it does so correctly.
We consider dishonest consumers later in Section \ref{sec:problems}.

For completeness, we identify the following additional attacks and show why 
we exclude them from the discussion.
\begin{itemize}
\item A router that (1) does not generate \pints\ when necessary, or (2) generates 
\pints\ without forwarding content downstream. Both cases can be reduced to packet loss 
events. We do not address these attacks since the router's malicious behavior 
cannot be detected.
\item A consumer that continuously generates interests to inflate accounting 
information. If aggregate or distinct accounting is required, the producer will 
not be able to detect such malicious behavior.

On the other hand, if individual accounting is required, consumer-specific data can be used to 
detect continuous requests. However, this scenario can be reduced to an Interest Flooding 
(IF) attack \cite{gasti2013and,compagno2013poseidon}, which is outside the scope of this paper.
A similar argument applies for distinct accounting information.
\item An external attacker that controls network links, that is capable of eavesdropping on, 
dropping, or replaying packets, including \pints. This attack is irrelevant if 
links are encrypted, which is a realistic assumption for adjacent routers. In most cases, consumers 
and producers are connected to edge routers using link-layer encryption, e.g., EAP~\cite{blunk1998rfc}.
\end{itemize}
%
Again, we do not consider an adversary that tries to inflate aggregate or distinct 
accounting information. As previously argued, this cannot be prevented 
deterministically due to the likely usage of multicast forwarding strategies. 
Therefore, we only consider security of correct individual accounting information. 
More formally, we define a secure accounting technique as follows.

\begin{definition}
\label{def:correctness_with_adv}
An accounting technique is {\em secure} with respect to \Adv\ if it is correct and all \Adv\ 
malicious behavior can be detected. 
\end{definition}

\subsection{Mitigating Forgeries and Replay Attacks} \label{sec:mitigating-replay}
Section~\ref{sec:pints} mentioned several options for generating consumer-specific 
data. However, in order to prevent inflation attacks, such data must be unforgeable 
and resistant to replay attacks. We define {\em secure} consumer-specific data as follows.
\begin{definition}
Consumer specific data is {\em secure} if it can be authenticated by at least the producer, 
and is neither forgeable nor subject to replay attacks.
\end{definition}
Providing replay resiliency can be accomplished if consumer-specific data carries a nonce $r$ and 
a timestamp $t$. If producers receive consumer-specific data with duplicate $r$ in the same 
time window to which $t$ belongs, such data is discarded. Using these values, 
secure consumer-specific data \secdata\ takes the following form.
\begin{align}
\label{equ:sec-crsd}
\secdatam = \Big[ &\cdatam || r || t, \Big. \nonumber\\
\Big. &\textsf{f}_{k} \left( \cdatam || r || t || Int.N \right) \Big]
\end{align}
where \cdata\ is consumer-specific data formed according to 
Section~\ref{sec:pints-format}\footnote{Note that \cdata\ and {\tt Cdata} are different. The former 
is generated by consumers and assigned to {\tt Payload} field of interest, while the latter is 
a field in a \pint\ and may contain none or many \cdatas.}, 
$\textsf{f}_{k}(\cdot)$ is a function that provide unforgeability, 
and $k$ is a secret key (either symmetric or private key.) 
Note that interest name in the $\textsf{f}_k(\cdot)$ computation is used to bind 
\secdata\ to the interest to which it is appended. This prevents \Adv\ from 
using the same \secdata\ for generating multiple \pints\ with different names.
The function $\textsf{f}_{k}(\cdot)$ can be a Hash Message Authentication Code 
(HMAC)~\cite{krawczyk1997rfc} (if consumers share secrets with producers) or
a digital signature generation function. Each method has well-known advantages 
and drawbacks. Symmetric HMAC-based functions are generally much less costly 
than digital signatures. However, the former requires {\em a priori} key distribution. 
On the other hand, using digital signatures involves in-line distribution and 
on-line verification of signers' public keys.
This method of generating \secdata\ incurs a lot of computational and storage overhead. 
In addition to signature verification, producers are required to maintain a list of all 
received nonces in the acceptable current time window for each accountable content.


At this point, it becomes clear that unforgeability and replay resiliency cannot be achieved unless 
{\em secure} consumer-specific data is used. This is not possible 
if aggregate or distinct accounting is required, where consumer-specific
data is not provided. One way to solve the problem is to include 
\secdata\ in all interests regardless of the accounting type required. This is 
impractical as it introduces unnecessary overhead for both consumers and producers.

\subsection{Preserving Consumer Anonymity}\label{sec:anonymous-id}
In order to provide {\em secure} individual accounting in the presence of \Adv, consumer-specific 
data should be generated securely. Digital signatures, by nature, reveal the consumer's identity, and
HMAC tags allow separate interests to be linked together since a key identifier must be included in the interest
to properly verify each HMAC. 
To this end, we develop a technique for generating \secdata\ anonymous to all network entities, except producers. 

We begin with the notion of consumer-specific data indistinguishability, which is necessary to maintain 
anonymity among an arbitrary set of consumers. 
\begin{definition}
Let $Cr^{a}$ and $Cr^{b}$ be two consumers generating two consumer-specific data values $\cdatam^{a}$ 
and $\cdatam^{b}$ in two different interests for the same content object $CO$.   
Let \Eve\ be any eavesdropper (except the producer publishing $CO$) and not directly connected to either 
$Cr^{a}$ or $Cr^{b}$. Let the event of \Eve\ successfully revealing the source of 
$\cdatam^{a}$ and $\cdatam^{b}$ be denoted as 
$\Evem_{\mathrm{rev}} \left( \cdatam^{a}, \cdatam^{b} \right) = 1$.
We claim that these two interests are indistinguishable if the probability 
of $\Evem_{\mathrm{rev}} \left( \cdatam^{a}, \cdatam^{b} \right) = 1$ is no better than a random guess. 
That is, 
\begin{align*}
\Pr \Big[ \Evem_{\mathrm{rev}} \left( \cdatam^{a}, \cdatam^{b} \right) = 1 \Big] \le \frac{1}{2} + \epsilon(n),
\end{align*}
for any negligible function $\epsilon$ and security parameter $n$.
\end{definition}
Moreover, we assume that consumers know the producer's public key $pk$ before 
requesting its content; see Section \ref{sec:problems} for justification. 
Let \adata\ denote an anonymous consumer-specific data of the form:
\begin{align}
\label{equ:a-crsd}
\adatam = \mathsf{Enc}_{pk} \left( \secdatam \right)
\end{align}
where $\mathsf{Enc}_{pk}(\cdot)$ is a public key encryption function using $pk$, and \secdata\ is 
formed according to Equation~\ref{equ:sec-crsd}.

To prevent \Eve\ from learning that two interests are generated by the same consumer, 
\adata\ values in the two interests should be indistinguishable. This can only be achieved if 
$\mathsf{Enc}_{pk}(\cdot)$ is a CPA-secure public key encryption scheme, i.e., secure against 
Chosen Plaintext Attacks~\cite{katz2007introduction}. In some encryption schemes, this is
done by mixing in a random number (nonce) with the plaintext before encryption.

\begin{theorem} \label{thm:indist}
Assume a CPA-secure public key encryption scheme $\mathsf{Enc}_{pk}(\cdot)$. An \adata\ composed 
according to Equation~\ref{equ:a-crsd} guarantees {\em indistinguishable} consumer-specific data 
with a negligible probability of nonce collision.
\end{theorem}

\begin{IEEEproof}
The proof of consumer-specific data indistinguishability follows from the proof of CPA-secure 
public key encryption scheme~\cite{katz2007introduction}. We only prove that \adata\ generation 
guarantees negligible probability of nonce collision.

We assume individual accounting, and that $f$ is the frequency in which consumers send 
interests to a specific producer during a specific time window $w$, where each consumer 
generates appropriate \adata\ values in the interests. Let the number of interests sent be 
$s = f \times w$. We claim that the probability 
of any two \adatas\ in the set $\{ \adatam_1, \dots, \adatam_s \}$ being derived from colliding nonces is 
negligible in $N$, the length of the nonce in bits. Let this collision event be denoted as 
$\mathsf{Col}(\adatam_i, \adatam_j)$ for $i \ne j$ and $1 \le i, j \le s$. The probability of this
event occurring can be calculated according to the birthday paradox as follows.
\begin{align}
\label{equ:collision_prob}
\Pr &\Big[ \mathsf{Col}(\adatam_i, \adatam_j) = 1\ ;\ i \ne j\ ,\ 1 \le i, j \le r \Big] \nonumber\\
&= 1 - \left( \frac{2^N}{2^N} \times \frac{2^N - 1}{2^N} \times \dots \times \frac{2^N - s + 1}{2^N} \right) \nonumber\\
&= 1 - \frac{2^N!}{\left( 2^N \right)^s \left( 2^N - s \right)!} \nonumber\\
&= 1 - \frac{s! \cdot {2^N \choose s}}{\left( 2^N \right)^s}
\end{align}
Note that Equation~\ref{equ:collision_prob} assumes that $s<2^N$; otherwise, the collision 
probability is equal to $1$ according to the Pigeonhole Principle.
\end{IEEEproof}
We note that public-key operations to generate and verify \adata\ values may be overly 
expensive for some consumers. In such scenarios, a potential optimization is to use 
symmetric encryption. Specifically, \adata\ values can be computed as 
$\mathsf{enc}_{k}(\secdatam)$, where $\mathsf{enc}_{k}(\cdot)$ is a CPA-secure symmetric 
encryption function and $k$ is a shared secret between the consumer and the producer. 
This, however, requires additional operations for creating and managing shared secrets. In order 
for producers to determine which shared secret to use for decrypting received \adatas, 
consumers should add a form of cleartext identifier, e.g., shared key tag. Although this 
violates interests indistinguishability, it does indeed preserve consumer anonymity.

\section{Individual Accounting in Practice} \label{sec:problems}
In this section we describe some challenges related to transparently collecting 
individual accounting information. We then outline recommendations for applications 
wishing to implement accounting in CCN.

\subsection{Individual Accounting Challenges}
Thus far, we made several assumptions about consumers
as far as individual accounting information conveyed to producers:
\begin{compactenum}
	\item Consumers know \emph{what} accounting information is needed for
	a desired content object.
	\item Consumers know the producer's public key $pk$ used to encrypt $\secdatam$
	sections.
	\item Consumers behave honestly, i.e., for content that requires individual accounting
	information in the $\cdatam$, consumers supply \emph{correct} required information.
\end{compactenum}
The first assumption seems to be particularly problematic, especially if a given 
consumer has no prior relationship with a specific producer. However, there are 
at least two simple ways for consumers to learn what a producer expects in an
interest. 

First, recall that adherence to sound trust management at the network layer \cite{ghali2014network}
requires the consumer to know, when requesting content (issuing an interest),
the public key of the producer. This, in turn, means that the consumer pre-fetched 
the producer's public key. It is easy to extend the producer's public
key certificate to include desired accounting requirements for constructing correct
interests for that producer and the namespace included in the certificate.

One natural alternative is for consumer to ``blindly'' issue a trial interest for some
random content in the namespace of a given producer. This interest might 
not abide by the producer's accounting rules. However, the producer could then
reply with its public key certificate that would include the producer's 
accounting requirements. 


Without performing one of the aforementioned techniques, a consumer
cannot be expected to provide specific information in an interest 
since it does not have any expectation of the form of this information. 
We consider possible consumer behavior for each accounting type below. 
\begin{compactitem}
	\item Aggregate: For this type of accounting information, the $\cdatam$ fields are empty
	and \pints\ only contain a count. Thus, consumers are not required to have any a priori
	information when constructing their original interest.
	\item Distinct: $\cdatam$ contains random nonces for each interest issued, and \pints\ 
	carry collections of these nonces from consumers to producers along with a total count. 
	The nonce does not require any application-specific input from the consumer to generate;
	it is simply a random string. Thus, in this case, consumers are also not required
	to have any a priori information.
	\item Individual: For this type of accounting information, $\cdatam$ contains a very specific
	piece of consumer-specific data (identifier) for each interest, and \pints\ propagate 
	these values to producers. To be useful for individual accounting, producers must be able
	to utilize the provided information in order to identify each consumer. Furthermore, 
	it will likely be the case that different producers require different identification
	information, both in its content and representation. Thus, consumers cannot be 
	expected to know \emph{which} type of consumer-specific data to provide in an interest
	without having been told beforehand.
\end{compactitem}
It is clear that individual accounting information necessitates some initial interaction
or registration phase, wherein consumers are given the interest $\cdatam$ requirements
and also the public key used when generating $\secdatam$ or $\adatam$ values in interests.
Note that the issue of public key identification is analogous to the problem of not knowing
\emph{what} consumer-specific data to use for a given interest.

This interaction or registration step covers assumptions (1) and (2) above, but it does not
address assumption (3). Namely, even if a consumer has all of the information at their disposal
needed to construct a valid interest for a given content object, what happens if they maliciously
choose to use the \emph{wrong} information? Such an adversary can easily obtain data from 
router caches without having to provide the correct accounting information to the producer, 
thereby effectively bypassing the accounting mechanism.

The core problem is that routers have no means to determine if the information contained
in interest $\cdatam$ fields is \emph{correct}\footnote{This means that the provided
information is the one required by the producer.} for a given content object. To be able
to verify this information routers would have to possess (or be told) some piece of 
information for each cached content that requires individual accounting. Only then can
routers verify $\cdatam$ field values before replying with a cached content. 
Not only is this unreasonable for routers, it also means that anonymous consumer-specific
data, as described in Section \ref{sec:anonymous-id}, is no longer feasible. 
Since routers must be able to verify the $\cdatam$ for each interest associated,
the consumer must necessarily reveal \emph{some} information about their identity.

Therefore, we claim that assumption (3) is not realistic at the network-layer in the presence
of caches and dishonest consumers. This means that individual accounting must be handled at the
application layer. The proof of this claim can be argued from the above discussion.

\subsection{Recommendations}
Given the previous discussion, we present some recommendations for collecting
accounting information in CCN. First, if individual accounting
information is needed, producers must simply set all content cache time to zero (0). 
This will force all interests to be routed to the producer without being
satisfied from an in-network cache. If an interest for content that requires individual
accounting is received and the required accounting information is missing, 
producers should reply with a Negative Acknowledgment (NACK)  
\cite{compagno2015nack} indicating consumer-specific data requirements
to obtain said content. The consumer can then re-issue an interest with the correct information. 
If consumers go through a preliminary registration step, this accounting
requirement information can be obtained once and then used for all subsequent 
interests, thereby removing the need for an additional round-trip.
Observe that since producers process all interests before responding with content, they
can determine if a given interest for individual accountable content is valid and thus
detect behavior by malicious dishonest consumer. 


For aggregate and distinct accountable content, consumers should always 
generate a random nonce and include it in $\cdatam$. If a router caches the content
and its {\tt ACCT} flag is {\tt AGGREGATE}, then $\cdatam$ values can simply be dropped 
when \pints\ are generated. Otherwise, if the router caches the content and its 
{\tt ACCT} flag is {\tt DISTINCT}, the nonce is copied into the generated \pint. 
This is a simple modification to the router \pint\ generation procedure described in 
Algorithm \ref{alg:pint-generation}, and induces no significant overhead for consumers
or routers. 

This simple policy can be extended to \emph{all interests}. Since consumers are not 
generally expected to know what type of accounting information is required for content, 
they can blindly generate a nonce for each interest they issue. CCN routers
will then correctly propagate these nonces in \pints\ to the producer according
to the rule above. As previously noted, NDN already supports default nonce generation in 
interests (but for the purposes of interest loop detection). The CCNx protocol needs to be 
updated to include this requirement.



\section{Analysis and Experimental Assessment} \label{sec:experiments}
In Section~\ref{sec:open-env}, we proposed two fundamental techniques for propagating accounting information to producers: encryption-based and \pint-based solutions. The former technique is beneficial in that it is entirely transparent to the routers. Encryption-based accounting, which is a form of access control, is an application-layer concern, and therefore the routers do not require any modification to support the scheme. Conversely, accounting based on \pints\ requires routers to execute the {\sf \pint-Generation} procedure upon every cache hit to generate \pints, and also forward \pints\ towards producers using the same data plane logic as normal interest messages.

\begin{figure}[t]
\centering
\includegraphics[width=\columnwidth]{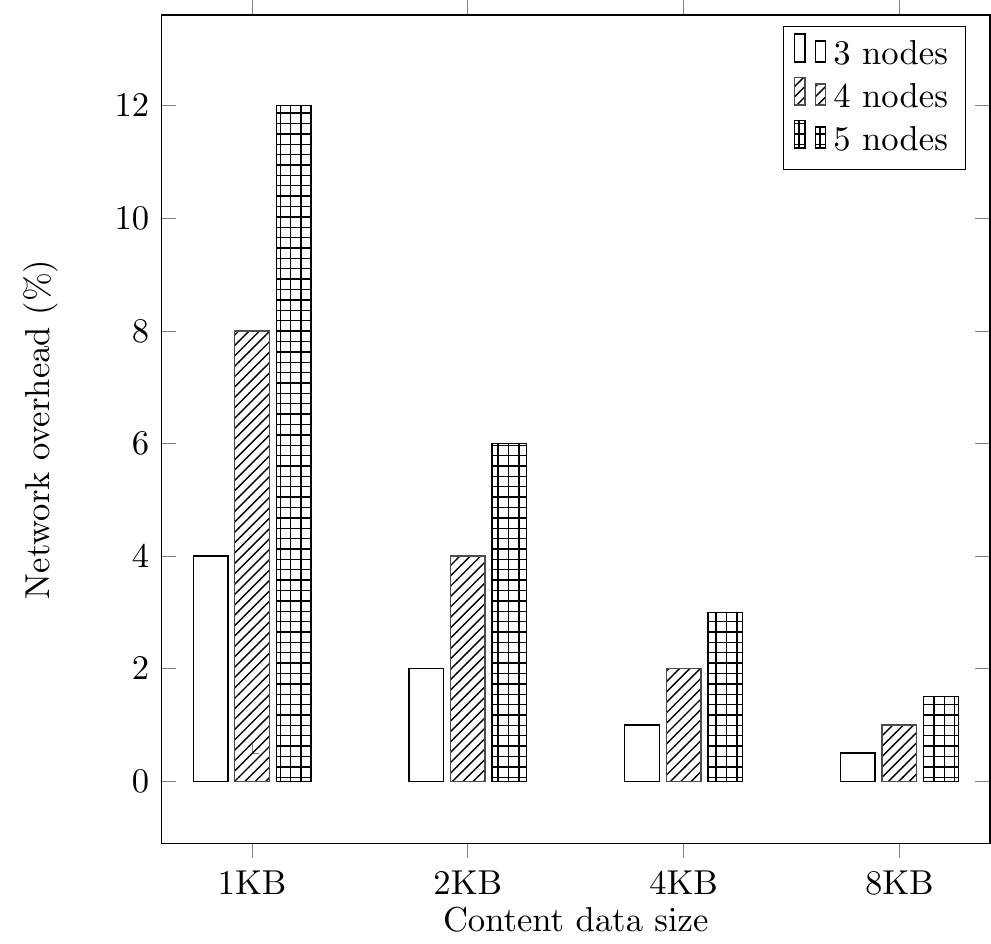}
\caption{Network overhead imposed by forwarding \pints.}
\label{fig:network_overhead}
\end{figure}

\begin{figure*}[t]
\begin{center}
\subfigure[Path topology with $5$ nodes (1 consumer, 3 routers, and 1 producer),\newline$A = 500$ and $M = 1000$.]
{
	\includegraphics[width=\columnwidth]{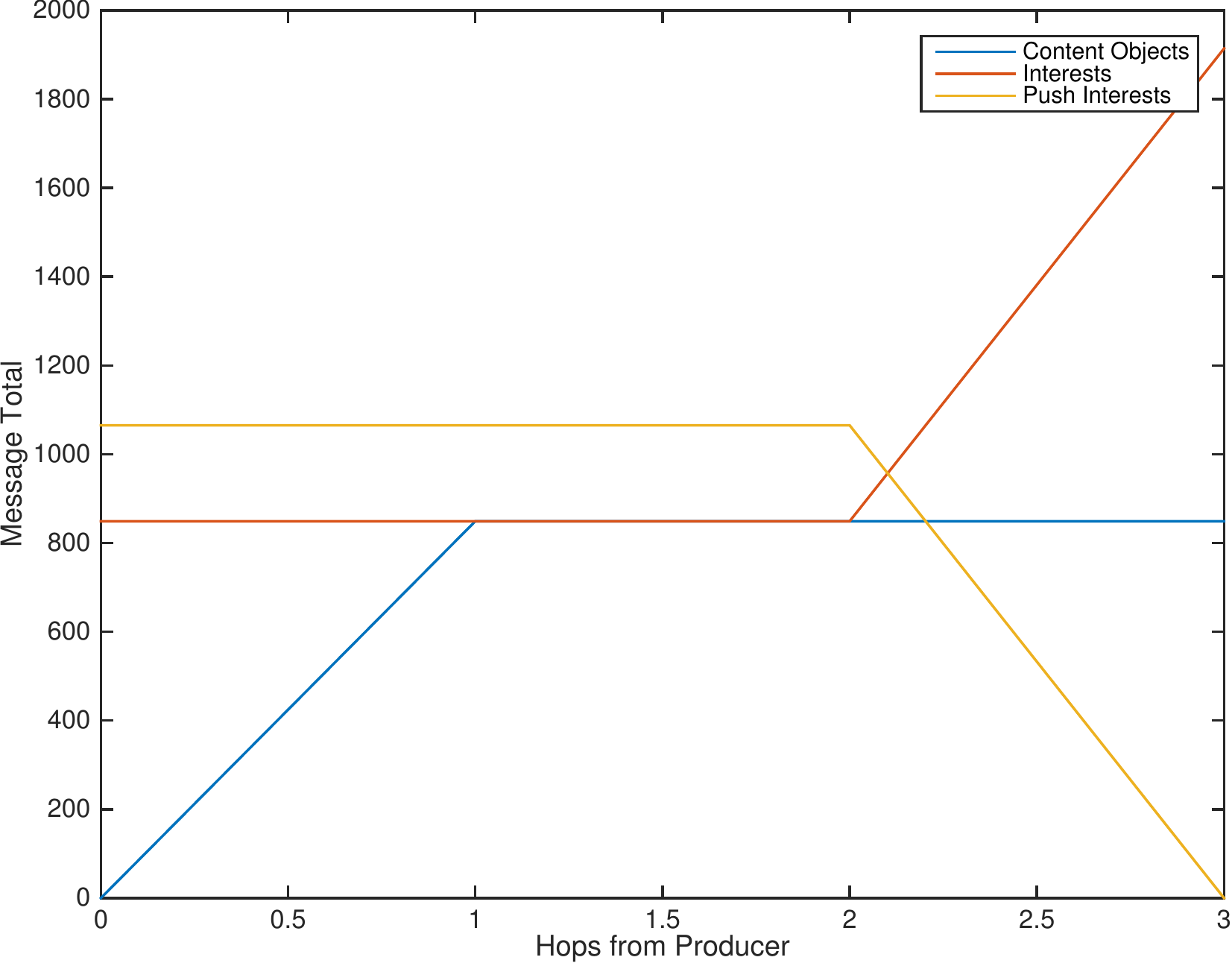}
	\label{fig:path_5_a500_m1000}
}
\subfigure[Path topology with $5$ nodes (1 consumer, 3 routers, and 1 producer),\newline$A = 500$ and $M = 10$.]
{
	\includegraphics[width=\columnwidth]{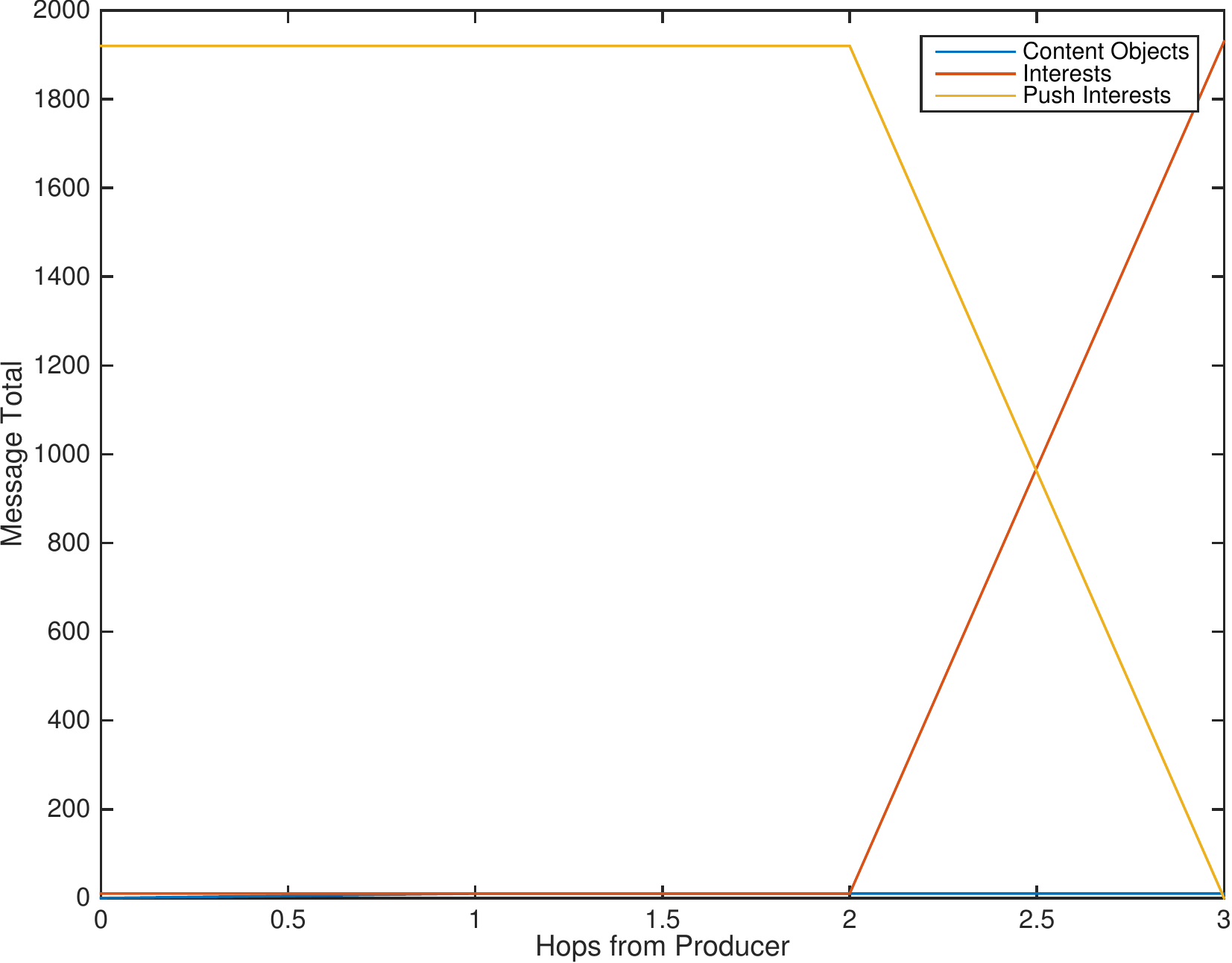}
	\label{fig:path_5_a500_m10}
}
\subfigure[Binary tree of height $5$ (32 consumers, 30 routers, and 1 producer),\newline$A = 500$ and $M = 1000$.]
{
	\includegraphics[width=\columnwidth]{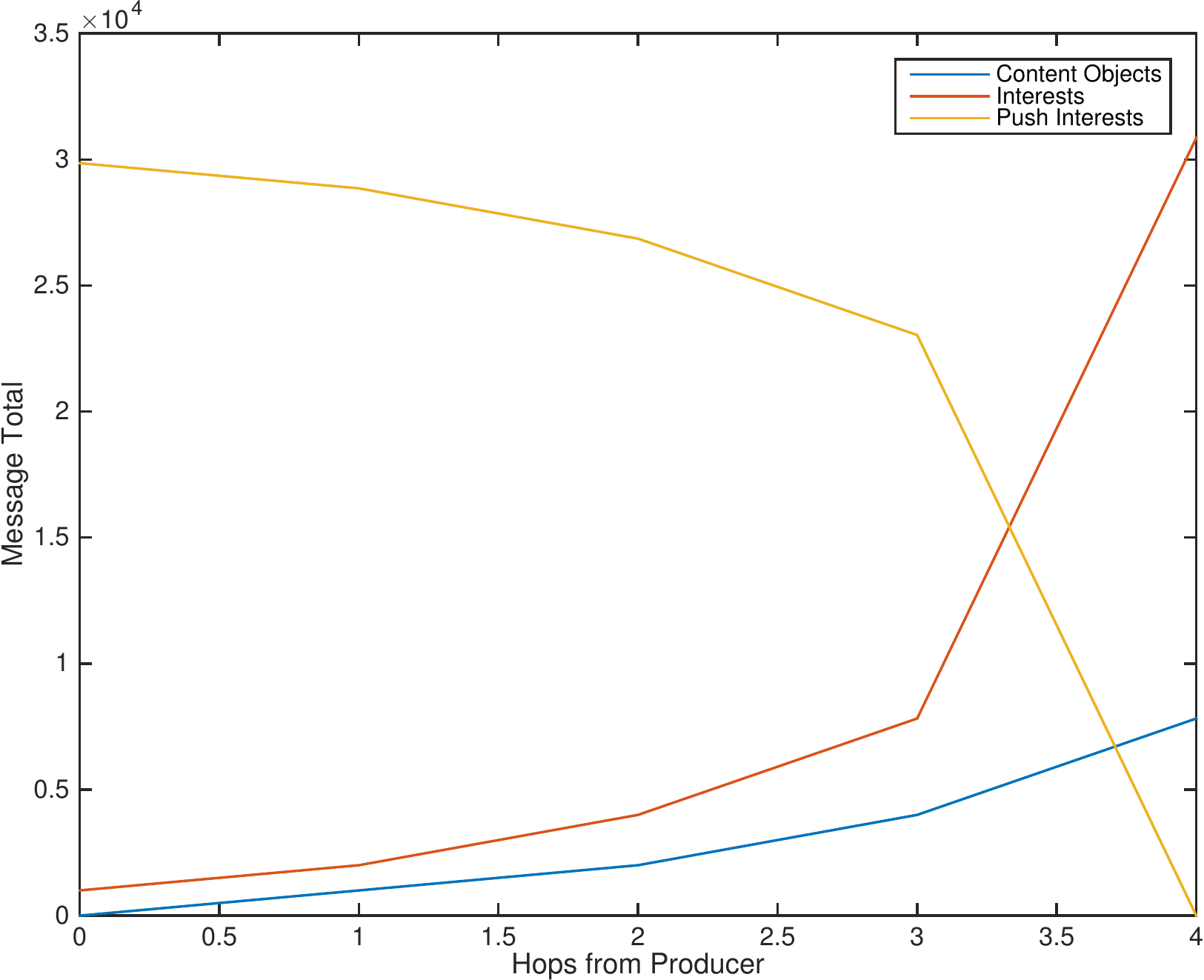}
	\label{fig:tree_2_5_a500_m1000}
}
\subfigure[Binary tree of height $5$ (32 consumers, 30 routers, and 1 producer),\newline$A = 500$ and $M = 10$.]
{
	\includegraphics[width=\columnwidth]{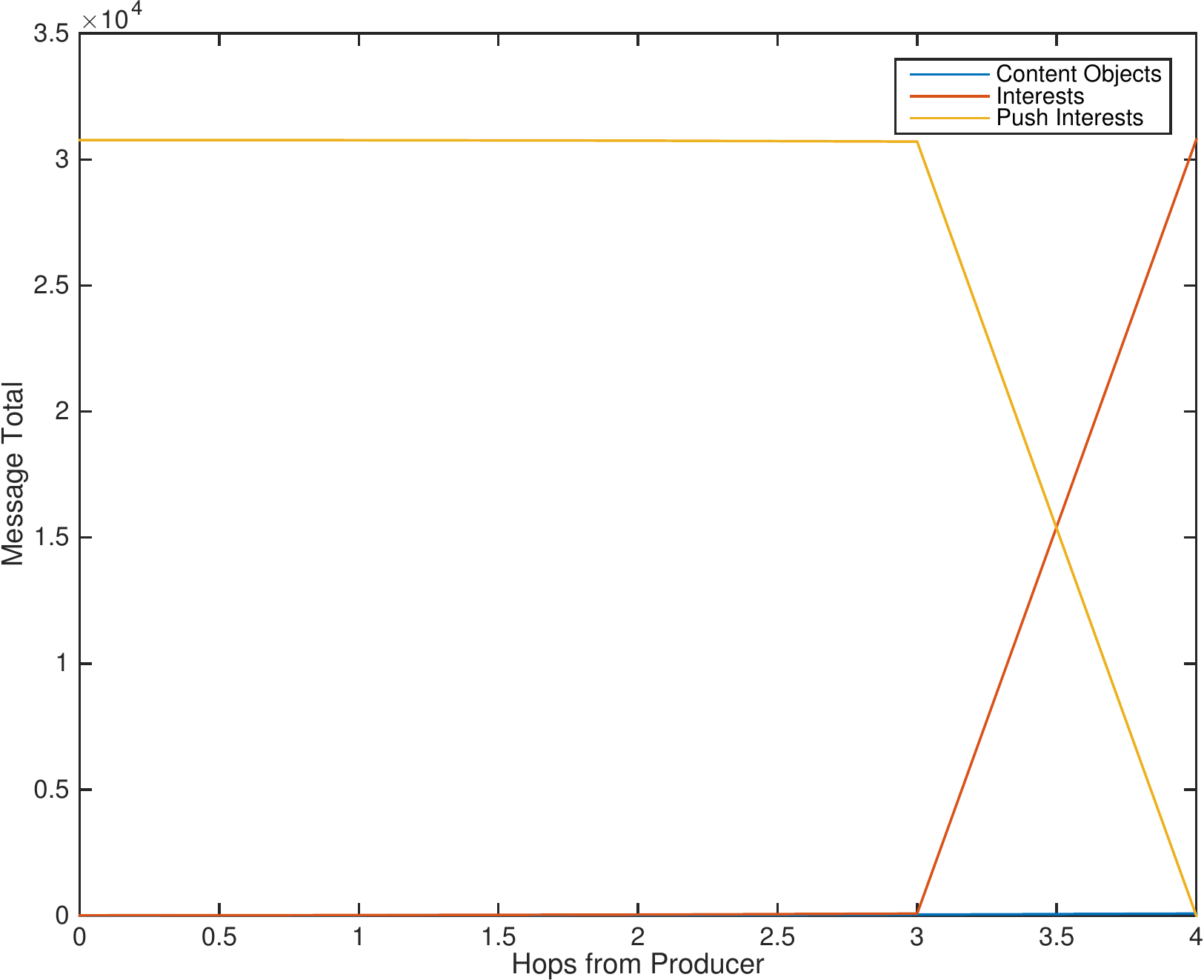}
	\label{fig:tree_2_5_a500_m10}
}
\caption{\pint-based accounting overhead in networks with path and tree topologies.}
\label{fig:accounting-results}
\end{center}
\end{figure*}

Consider a scenario in which we have $k$ consumers $Cr_1,\dots,Cr_k$ and a single producer $P$.
Let $Cr_i,R_1,\dots,R_l,P$ be a consumer-to-producer path traversed by interests issued by consumer $Cr_i$
for the accountable content object $CO$. Let $R_c$, $1 \leq c \leq l$ be the router at which $CO$ is cached.
Furthermore, let $p_l$ be the number of messages traversing the $R_1-R_c$ path in one direction, and let $p_r$
be the number of messages traversing the $R_c-P$ path in one direction. Finally, let $\gamma$ be the number
of content requests issued by all consumers $Cr_i$, $i=1,\dots,k$, along the $R_1-P$ path. Recall that
encryption-based accounting requires consumers to issue at least two (2) interests to access accountable
content: one interest is issued for the content itself, and then at least one more is issued to request
the corresponding decryption keys. The former interest will traverse the $R_1-R_c$ path, whereas all decryption
key interests will traverse the full $R_1-P$ path. Thus, in this case, $p_l = 4\gamma$ and $p_r = 2\gamma$.
Conversely, in the \pint-based approach, a single interest is issued for $CO$ on the $Cr-R_c$ path, and
then a \pint\ is generated at $R_c$ and forwarded along the $R_c-P$ path. Therefore, in this case, $p_l = 2\gamma$
and $p_r = \gamma$. Note that the case where $R_c = P$ is identical to the scenario where there are no
in-network caches, in which case there would be no \pints\ generated. This case performs worse than
the \pint-based variant since $p_l = 2\gamma$ and $p_r = 2\gamma$ (there is a single RTT from the $Cr_i$
to $P$ for $CO$).

Notice that the differences in $p_l$ are due to the fact that, unlike interests, \pints\ have no response from
the producer. In fact, the \emph{network overhead}, in terms of the number of messages, of the encryption-based
accounting solution is exactly twice that of the \pint-based solution, and the network overhead of the
cacheless variant (which, again, invalidates the need for \pints\ and accounting information) is more
than the \pint-based solution as well. Furthermore, producers, and consumers incur additional overhead since
encryption and decryption must be performed, respectively, in order to consume content. Therefore, in our initial
experimental assessment, we limit our focus to the \pint-based accounting solution, since we feel that it is (a) a more
efficient technique and (b) also proportional to the overhead incurred by network entities in the encryption-based
scenario. In our experimental analysis, we assume that adding the \pint\ generation procedure is a constant time
operation for routers. We also assume that any shared-secret management protocols, such as those that might be used
to establish or acquire a shared HMAC key, are done offline and are therefore not part of the real-time or online communication.

If interests are satisfied from the cache of $R_c$, all upstream routers on the consumer-to-producer path incur the
overhead of forwarding \pints\ to the producer. Figure~\ref{fig:network_overhead} shows this overhead as a function
of corresponding content size and the number of links between the router generating \pint\ and the producer. It is
calculated as the the ratio of the extra bytes (due to the forwarded \pints) traversing each link over the size of
the corresponding content object. The $x$-axis represents content data size, without including the header. We
calculated the overhead in three line topologies containing 3, 4, and 5 nodes consisting of 2, 3, and 4 links
respectively. The first node is the consumer $Cr$ and the last node is the producer $P$. For the purpose of this
exercise, we assume the following:
\begin{figure}
\center
\includegraphics[width=0.86\columnwidth]{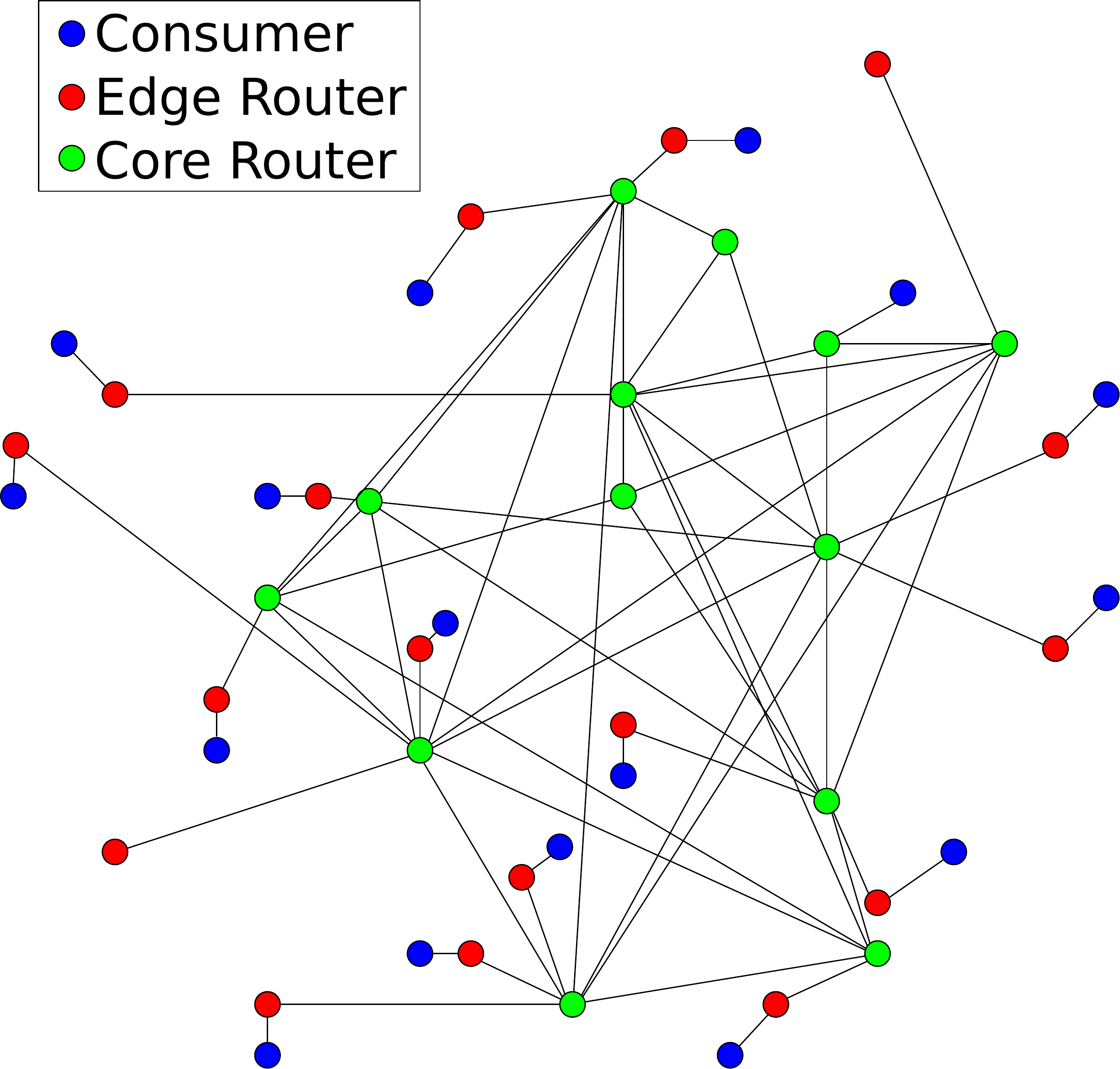}
\caption{DFN topology - each edge router above is connected to a group of NDN consumers.}
\label{fig:dfn_topology}
\end{figure}
\begin{figure}
\center
\includegraphics[width=\columnwidth]{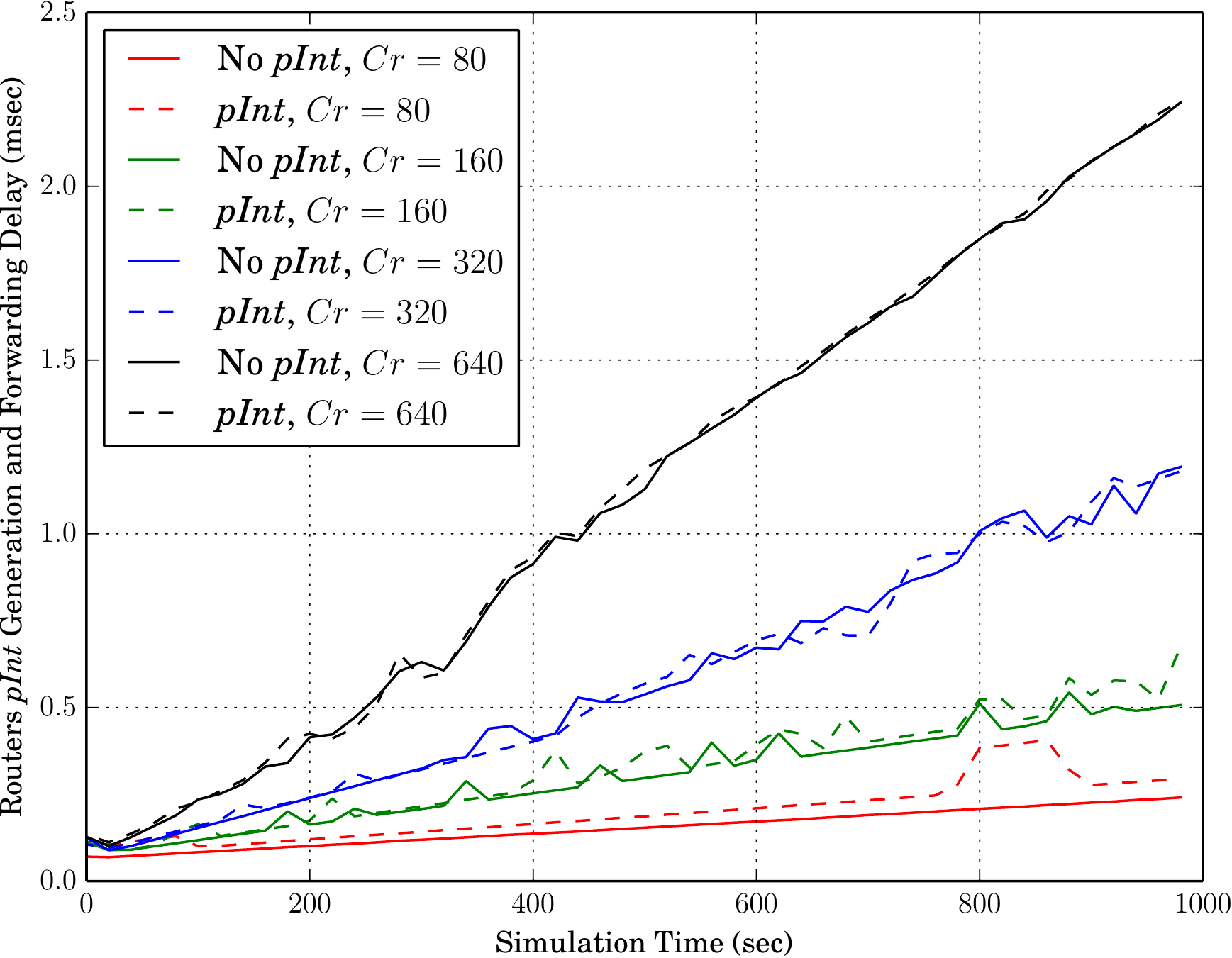}
\caption{DFN topology - \pints\ generation overhead at routers.}
\label{fig:dfn-pint-generation-overhead-delay}
\end{figure}
\begin{itemize}
\item Any content requested by $Cr$ can always be satisfied by the first hop (consumer facing) router's ($R$) cache, i.e., cache hit rate is $100\%$. This accounts for the highest network overhead since \pint\ must traverse all links except the first one connecting $Cr$ with $R$.
\item Each router's FIB is pre-configured to forward all interests and \pints\ towards $P$.
\item Interest, \pint, and content object headers only contain a name of length $40$B.
\end{itemize}

The results from this experiment show that, as the size of content objects grow, the bandwidth overhead induced
by \pints\ decreases. This overhead would increase if more complex topologies were considered, e.g., $k$-ary
trees rooted at $P$. However, it would see the same decline as the content object size increased.

\subsection{Message Count Overhead}

\begin{figure}
\center
\includegraphics[width=\columnwidth]{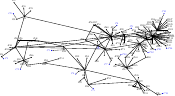}
\caption{AT\&T topology - each edge router above is connected to a group NDN consumers.}
\label{fig:att_topology}
\end{figure}

\begin{figure}
\center
\includegraphics[width=\columnwidth]{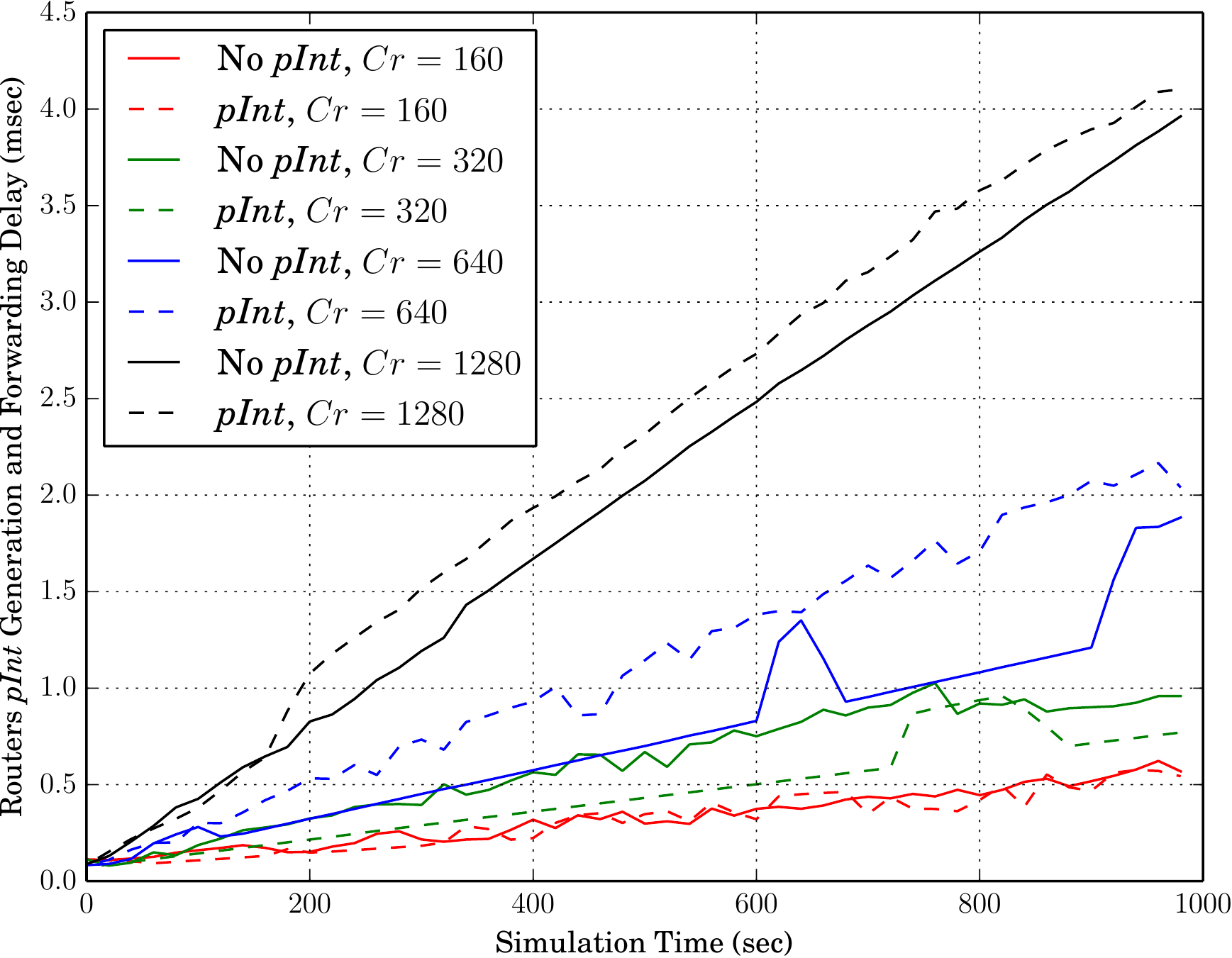}
\caption{AT\&T topology - \pints\ generation overhead at routers.}
\label{fig:att-pint-generation-overhead-delay}
\end{figure}

To further understand the performance impact of \pint-based accounting in different topologies, we studied the amount of
overhead incurred by each entity -- consumer, router, and producer -- in the network as a function of the distance from
the producer. To do this, we implemented a custom discrete-time event-driven simulator that models a variety of CCN network
topologies: paths and binary trees. Consumers are configured to issue interests for a single producer at a Poisson rate with parameter $A$. The names of each interest are uniformly sampled from a pool of $M$ names. Each router invokes the {\sf \pint-Generation} procedure upon every cache hit. We place no restriction on cache sizes, since the set of possible content objects is small enough to fit within any reasonable sized cache. Specifically, producers respond to interests with content objects with a fixed payload size of $1$MB.

We argue that the number of messages is indicative of the bandwidth overhead induced by \pints. This is because the size of \pints\ will be proportional to the size of interests, even with {\em secure} consumer-specific data. Therefore, it is not the size of these messages that is important -- it is the sheer quantity which are processed by network elements. Thus, we are concerned with the number of messages that propagate from routers to the producer.

We study this overhead in networks with path and tree topologies. For simplicity, we restrict our analysis to
5-hop paths and binary trees of height 5. By varying $A$ and $M$, we show how many messages of each
type are processed by each entity as a function of the distance from the producer.
Figure \ref{fig:accounting-results} illustrates the obtained results. Clearly, as $M$ decreases, the
likelihood of cache hits increases. This results in a clear increase in \pint\ processing at each
upstream entity from the cache hit location. For example, in the cases where $M = 10$,
approximately $99\%$ of all messages processed by routers upstream of cache locations were \pints.

The interest request rate is highly dependent on the type of application. High request rates for popular
content, which is likely to be cached, will lead to a proportionally high number of \pints\ propagating
upstream to the producer. If interests are issued for unpopular or uncached content, then approximately
the same number of interests will be propagated upstream. In other words, from the perspective of the
producer, the sum of the received interests and \pints\ will be equal the total number of content
requests from all consumers: {\bf the producer overhead is linear in the number of content requests.}
The difference in these two cases is that \pints\ are typically smaller in size than their interest counterparts.

\subsection{Router Overhead Assessment}
To measure router overhead incurred by generating and forwarding \pints\ we extended
ndnSIM \cite{afanasyev2012ndnsim}, a simplified implementation of NDN architecture as a NS-3
\cite{ns3} module for simulation purposes, to support \pints. With this modified
architecture, we two sets of experiments using two different topologies:
\begin{itemize}
\item The DFN network, Deutsches ForschungsNetz (German Research Network) \cite{DFNverein, DFN-NOC}:
a German network developed for research and education purposes which consists of several
connected routers positioned in different areas of the country, as shown in Figure
\ref{fig:dfn_topology}. The network consists of a total of 30 routers. Blue dots in the
figure represent group of consumers connected to edge routers (red dots), while green dots
represent core network routers.
\item The AT\&T backbone network~\cite{compagno2013poseidon}: shown in Figure
\ref{fig:att_topology}, this network consists of more than 130 routers, and each
logical consumer in the figure represents multiple physical consumers connected to an edge router.
\end{itemize}
In all of our experiments, consumers issue interests at a rate of $10$ interest per second
for the same content with the name \ccnname{/prefix/A/00}. To capture the worst case scenario,
wherein the maximum number of \pints\ are generated, we (1) disable interest collapsing,
and (2) set the \texttt{ExpiryTime} of the request content to be equal to the simulation time,
ensuring that this content is cached at routers throughout the whole duration of the simulation.
This forces routers to generate a single \pint\ for every cache hit, resulting in the maximum amount
of \pints\ that can be possibly generated.

We measure the overhead required by routers to generate and forward \pints\ as compared
the the case where \pints\ are not generated. Figure \ref{fig:dfn-pint-generation-overhead-delay}
shows the router overhead in the DFN topology parameterized by the number of consumers
connected to edge routers ($80$, $160$, $320$, and $640$). We observed that even with $640$ consumers in
the network, the overhead of an average router when generating \pints\ is negligible. Similarly, Figure
\ref{fig:att-pint-generation-overhead-delay} shows the overhead of generating and forwarding
\pints\ by routers in the AT\&T topology. In the case with $1280$ consumers, routers experience
a $15\%$ additional overhead while generating \pints, which we consider to be negligible and
a difference that can be recovered with better routing hardware.

\section{Related Work} \label{sec:related}
Network-layer accounting in CCN and related interest-based ICN architectures remains
an open topic in the literature \cite{xylomenos2014survey}. However, certain economic aspects,
such as \emph{how} to set and enforce prices, has been widely discussed
\cite{araldo2014design,pham2013pricing}. These results imply an application-layer strategy whereby
payment (not usage) information is willingly sent on behalf of the consumer. This conflicts with
the approach advocated by Agyapong et al. \cite{agyapong2012economic}, wherein only
ISP-related entities (i.e., not consumers or producers) are involved in payment coordination.
\cite{agyapong2012economic} considers producer payment as an application layer concern.
Our accounting techniques facilitates a blend between these two schemes wherein usage
and payment information are sent \emph{autonomously} on behalf of the network-layer for
consumers (end-hosts) and routers. ISP entities and producers are informed of
usage information for billing purposes, and can follow up with payment collection
at a later point in time. 

Patan\'{e} et al. \cite{patane2014economics} study a similar problem in the context of IP-compatible
architectures. Specifically, they focus on ones with dedicated router caches
like Content Distribution Networks (CDNs) and
transit networks that chauffeur traffic between different ISP provider networks.
Payment policies proposed in \cite{patane2014economics} are identical, though.
All parties pay for the resources which were used to deliver their content.
Patan\'{e} et al. also neglect to discuss means by which this payment and usage information can be
propagated. Similar to \cite{pham2013pricing}, Kocak et al. \cite{kocak2013effect} discuss methods
where content providers can coordinate price information and contracts between ISPs.
Kocak et al. also opt for an open, unfederated approach, which fits with our model of autonomous
accounting information propagation.

Another important element of this work is the generation of secure consumer-specific data in \pints.
There is rich literature of packet-level authentication in the IP-based Internet, much of which
is contained in \cite{lagutin2008redesigning}. However, techniques such as digital signatures
and symmetric-key MACs require some possibly unrealistic assumptions, such as shared
keys amongst all pairs of routers and trusted third parties for key generation and management.
Using improved public-key cryptographic algorithms based on elliptic curves can help improve
the signature scheme efficiency \cite{johnson1998elliptic}, as with DNSCurve
\cite{bernstein2009dnscurve}. However, the sheer volume of interests in CCN and related
ICNs will very likely be substantially larger than DNS queries in IP networks, leading to
only relatively modest improvements in performance.

\section{Conclusion}
This paper represents the first attempt to address accounting in CCN. 
It presented a simple and lightweight accounting technique and showed 
how to enhance it with security without significant burden to consumers, producers, 
and routers. We analyzed performance of the proposed technique and 
demonstrated that secure accounting is both possible and practical in CCN. 

\balance

\bibliographystyle{IEEEtran}
\bibliography{IEEEabrv,references}

\end{document}